\documentclass[prd,amsmath,amssymb]{revtex4}
\usepackage{graphicx}
\usepackage{color}
\usepackage[colorlinks=true,linkcolor=red]{hyperref}
\usepackage{hyperref}

\begin{document}
\title{The general relativistic infinite plane}
\author{Preston Jones}
\email{prjones@ulm.edu}
\affiliation{Department of Mathematics and Physics, University of Louisiana at Monroe,
Monroe, Louisiana 71209-0575 \\}
\author{Gerardo Mu{\~n}oz}
\email{gerardom@csufresno.edu}
\author{Michael Ragsdale}
\email{raggy65@csufresno.edu}
\author{Douglas Singleton}
\email{dougs@csufresno.edu} 
\affiliation{Physics Dept., CSU Fresno, Fresno, CA 93740-8031}

\date{\today}

\begin{abstract}
Uniform fields are one of the simplest and most pedagogically useful 
examples in introductory courses on electrostatics or Newtonian 
gravity. In general relativity there have been several proposals as to what 
constitutes a uniform field. In this article we examine two metrics that 
can be considered the general relativistic version of the infinite 
plane with finite mass per unit area. The first
metric is the 4D version of the 5D ``brane" world models which are the starting
point for many current research papers. The second case is the cosmological domain 
wall metric. We examine to what extent these different metrics match or deviate 
from our Newtonian intuition about the gravitational field of an
infinite plane. These solutions provide the beginning student in general
relativity both computational practice and conceptual insight into Einstein's
field equations. In addition they do this by introducing the student 
to material that is at the forefront of current research.
\end{abstract}

\maketitle

%%%%%%%%%%%%%%%%%%%%%%%%%%%%%%%%%%%%%%%%%%%%%%%%%%%

\section{Introduction}

The homogeneous gravitational field (\textit{i.e.} the $\vec{g}$ in 
${\vec{F}}=m{\vec{g}}$) is one of the most commonly used examples in introductory
mechanics courses. Students later find out that this is an approximation
used when the observer is small compared to the radius of the spherical
gravitating body on which they reside -- if $d$ is the size of the observer
and $R$ the radius of the spherical body then $d\ll R$. In Newtonian gravity
a truly uniform field would be produced by an infinite plane with area mass
density $\sigma $. Infinite planes of charge are
also among the first examples students encounter in introductory
electrostatics. Using the close connection between Newtonian gravity and
electrostatics one can find the gravitational potential for such an infinite
mass plane. The Newtonian gravitational potential, $\phi $, for such a source
is given by

\begin{equation}
\vec{g}=-\nabla \phi =-2\pi G\sigma  [\Theta (z)-\Theta (-z)] \widehat{z}
\qquad \rightarrow \qquad \nabla ^{2}\phi =4\pi G\sigma \delta (z) ,
\label{laplace}
\end{equation}
where $\Theta (u)$ is the step function ($%
\Theta (u)=0$ for $u<0$ and $\Theta (u)=1$ for $u\ge 0$); $\delta (z)$ is
the Dirac delta-function which results from differentiating the step
function. The last equation is obtained by taking the divergence
of the first equation. The solution, $\phi (z)$ to \eqref{laplace} is 
$\phi (z)=2 \pi G\sigma |z|$.

We want to construct the general relativistic version of the infinite
plane. We find that there are two metrics which have some (but not all)
of the characteristics of the Newtonian infinite plane.
These two metrics are excellent examples to introduce beginning
students to both calculational and conceptual aspects
of general relativity. In addition one of the two examples 
has connections with current
research in ``brane" world models. This can help generate interest in beginning
students of general relativity, by showing them they are not (in all cases)
far removed from frontier research. 

There have been many previous claims for constructing the general
relativistic field for an infinite plane (a nice discussion and
an extensive list of references on this subject can be found in \cite{gron}).
However, none of these solutions has the correct matter configuration for an
infinite plane, namely $\rho (z)\propto \delta (z)$ as in \eqref{laplace}.
In addition, one might expect that the general relativistic infinite plane would
in some sense give rise to a ``uniform'' gravitational field. We will
indeed find this to be the case since our first solution is a 4D
version of the 2D ``uniform" gravitational field introduced by Desloge \cite{desloge}. There
are, however, some important differences between the general relativistic
infinite plane and the Newtonian infinite plane which arise from the fact
that the field equations of general relativity imply the equations of
motion. Thus one cannot put down an infinite plane and expect it to
be static. We will find that in order to stabilize the general relativistic
infinite plane one needs to introduce pressures/tensions and, in one
case, a cosmological constant term.

In the end we find that the first solution is a 4D
version of the 5D ``brane world" models \cite{Gog1} \cite{Gog2} \cite{RS} which
have provided a new solution to the hierarchy problem (i.e., the question or puzzle as 
to why the gravitational interaction is many orders of magnitude weaker than 
the three other known interactions) as well as other open questions in particle physics and cosmology.
Reference \cite{rubakov} gives an excellent introduction to the topic of
5D and higher dimensional brane world models, their utility in solving
the hierarchy problem, and other open questions in particle physics and
cosmology that can be addressed by these models. The second solution is the
domain wall solution first given in \cite{taub} and studied in some detail in
\cite{vilenkin} \cite{ipser}. It is a simple solution which can again serve as an
excellent calculational exercise for beginning students. Domain walls are 
not phenomenologically viable, but they are often grouped together with cosmic strings, which are being actively studied. As such the domain
wall solution might provide a starting point for discussing exotic cosmological solutions
\cite{ajp-string}. In the concluding section we will give some discussion of the
physical meaning of the brane world and domain wall metrics, and give some
ideas for additional investigations one could assign as projects for students.

\section{General Relativistic Infinite Plane}

Starting with the Newtonian infinite plane we want to motivate the general
relativistic infinite plane. In the weak field limit ($GM/rc^{2}\ll 1$, where 
$M$ is the mass of the gravitating object and $r$ is some distance scale
involved in the problem) the relationship between the Newtonian potential
and the $g_{00}$ component of the metric is $g_{00}=-(1+2\phi/c^2 ) \rightarrow
-(1+2\phi)$ (from now on we set $c=1$). Thus, given that the 
Newtonian potential for the infinite plane is $\phi (z)=2 \pi G\sigma
|z|\equiv g|z|$ one might be tempted to try the metric $g_{00}=-(1+2g|z|)$, $%
g_{ii}=+1,$ where $i=,x,y,z$ and we have a gravitational acceleration $%
g=2 \pi G\sigma.$ One can easily check that this does not work, since for this metric 
only the $G_{xx}$ and $G_{yy}$ components of the Einstein tensor are non-zero.
Since $G_{\mu \nu }\propto T_{\mu \nu }$ this implies only the $T_{xx}$ and 
$T_{yy}$ components of the energy-momentum tensor are non-zero, whereas we
wanted only $T_{00}\propto \rho $ to be non-zero. We will find that it is not possible 
to construct a static mass distribution without pressures/tensions or possibly a cosmological constant.

One may notice that the trial metric -- $g_{00}=-(1+2g|z|)$, $g_{ii}=+1$ --
can be seen as some limit of the Rindler metric \cite{rindler} which is
Minkowski spacetime as seen by an observer undergoing constant, linear
acceleration,

\begin{equation}
ds^{2}=-\left( 1+gz\right) ^{2}dt^{2}+dx^{2}+dy^{2}+dz^{2}\approx -\left(
1+2gz\right) dt^{2}+dx^{2}+dy^{2}+dz^{2},  \label{rindler}
\end{equation}
where the last approximation is $gz\ll 1$. If we ignore the lack of absolute
value around $z$ this is just the trial metric. However, it can be shown 
\cite{desloge} \cite{tilbrook} that the Rindler metric is not a
solution to the Einstein field equations for any ponderable source of the
gravitational field, i.e., it is a vacuum spacetime. This can be seen
directly since the Rindler metric \eqref{rindler} can be obtained from Minkowski
spacetime via the transformation $t^{\prime }=[(1/a)+z]\sinh (at)$ and $%
z^{\prime }=[(1/a)+z]\cosh (at)$.

Continuing our search of the general relativistic analog of the infinite
plane we note that in \cite{desloge} the $g_{00}$ component for the uniform
gravitation field metric was given by $g_{00}=e^{2gz}$ (in \cite{desloge}
the coordinate was $x$ and $|g|=1$). Also noting that $e^{2gz}\approx 1+2gz$ for $%
gz\ll 1$ we might try using $e^{2g|z|}$ for $g_{00}$. This same association
between the approximate ($1+2 g z$) and exact ($e^{2gz}$) time component of the metric was
given originally by Einstein \cite{einstein11}. What, if anything,
do we do for $g_{ii}$ where $i=x,y,z$? If we make the simplest guess that
only $g_{00}$ should be non-trivial we arrive at

\begin{equation}
ds^{2}=-e^{2g|z|}dt^{2}+dx^{2}+dy^{2}+dz^{2}.  \label{visser}
\end{equation}
This is a 4D version of the exotic Kaluza-Klein metric proposed by Visser 
\cite{visser} which is considered a precursor to the current higher
dimensional brane world models. Again, if one calculates the Einstein
tensor $G_{\mu \nu }$, the only non-zero terms are
\begin{equation}
\label{visser2}
 G_{xx}=G_{yy} = \frac{\left(e^{g|z|} \right) ''}{e^{g|z|}} = g \delta (z) + g^2 ,
\end{equation} 
where the primes indicate differentiation with respect to $z$.
The $\delta (z)$ function appears because we are taking the 
second derivative of $|z|$.
Adding a cosmological constant term to the field equations would give non-zero values of 
$T_{00}$ and $T_{zz}$, but not of the form $\delta (z)$ needed for the mass distribution for an infinite plane.

To get rid of the $G_{xx} , G_{yy}$ terms one should take $g_{xx}, g_{yy}$
as non-trivial. The simplest guess -- that one should extend the ``warp"
factor $e^{2 g |z |}$ to encompass $dx^2 + dy^2$ -- turns out to be the correct
one. The metric now becomes

\begin{equation}  \label{gog}
ds^{2}=e^{2 g |z |}( - dt^{2}+dx^2+dy^2) +dz^2.
\end{equation}
The components of the Einstein tensor, $G_{\mu \nu}$, for \eqref{gog}
can be calculated by hand. This provides a good exercise for a beginning 
general relativity student, which is non-trivial, yet simpler than the 
Schwarzschild metric. The non-zero components are

\begin{eqnarray}
G_{xx} &=& G_{yy} = -G_{00} = \left( ( e^{g|z|} )' \right) ^2 + 2 e^{g|z|}( e^{g|z|} )''
= 3g^{2}e^{2 g|z|}+2g\delta (z)  \nonumber
\label{gmunu} \\
G_{zz} &=& 3 \left( \frac{( e^{g|z|} )'}{e^{g|z|}} \right) ^2
= 3g^{2}.
\end{eqnarray}
Now we finally see the presence of the $T_{00}\propto \delta (z)$
matter source implied by its presence in $G_{00}$. However, there are also
pressures in the $x, y$ directions since $G_{xx} = G_{yy} \ne 0 \rightarrow 
T_{xx} = T_{yy} \ne 0$, and the components 
$G_{00},G_{xx},G_{yy}$ have an extra, non-$\delta$ function term -- $\pm 3g^{2}e^{2 g|z|}$.
Finally, $G_{zz}$ is non-zero and equal to $3g^{2}$. These extra, non-$\delta$ function 
terms can be eliminated by the introduction of a cosmological constant term as follows

\begin{equation}
G_{\mu \nu }=8\pi GT_{\mu \nu } - 8\pi G\lambda g_{\mu \nu }\qquad \mathrm{where}
\qquad \lambda =-\frac{3g^{2}}{8\pi G}  \label{cc}
\end{equation}
As can easily be checked, this is exactly the 4D version of the 5D brane
models in \cite{Gog1} \cite{Gog2} \cite{RS}. The energy-momentum tensor for
our system can be read off by comparing \eqref{gmunu} and \eqref{cc} with the result 
\begin{equation}
T_{00} = - \frac{g}{4\pi G}\delta (z) ~, \qquad T_{xx}= T_{yy}=\frac{g}{4\pi G}\delta (z)  \label{em-tensor}.
\end{equation}
All other components are zero. Note that this is not
exactly the same as the Newtonian case of the infinite plane where 
$T_{00}\propto \delta (z)$ and all other components of $T_{\mu \nu }$ zero.
In the general relativistic case there are pressures in the $x$ and $y$
directions (the non-zero $T_{xx},T_{yy}$) as well as a negative cosmological
constant which is proportional to the square of the acceleration, $g^{2}$, and
thus to the square of the area density, $\sigma ^{2}$. 

There is another plane solution to the Einstein field equations which has
$T_{00}\propto \delta (z)$, and pressures in the perpendicular directions,
but without a cosmological constant. It is the 
domain wall metric first given by Taub \cite{taub} and investigated further
in \cite{vilenkin}. The domain wall metric is 
\begin{equation}
\label{domain-wall}
ds^2 = \left( 1 - 2 g | z | \right) ^{-1/2} (-dt^2 + dz^2)
+(1 - 2 g |z | ) (dx^2 +dy^2) ~,
\end{equation}
where as before $g = 2 \pi G \sigma$. Calculating the components 
of $G_{\mu \nu}$ for this metric again provides a simple calculational
exercise for beginning students. The only non-zero components are
\begin{eqnarray}
\label{domain-gun}
G_{00} &=& - \frac{\left( 1 - 2 g | z | \right)''}{\left( 1 - 2 g | z | \right)} = +2 g \delta (z) \nonumber \\
G_{xx} &=& G_{yy} = \frac{1}{4} \left( 1 - 2 g | z | \right) ^{1/2} \left( 1 - 2 g | z | \right) '' =
-\frac{1}{2} g \delta (z)~.
\end{eqnarray}
Using \eqref{domain-gun} in \eqref{cc} but with $\lambda =0$ the non-zero components of the
energy-momentum tensor are
\begin{equation}
T_{00} = \frac{g}{4\pi G}\delta (z)  ~, \qquad T_{xx}= T_{yy}= - \frac{g}{16\pi G}\delta (z)  
\label{em-tensor2}.
\end{equation}
Since the domain wall solution given in \eqref{domain-gun} \eqref{em-tensor2} has no 
cosmological constant one might prefer this as the general relativistic version
of the infinite plane. However, from \eqref{domain-gun} one can see that the domain 
wall solution has a singularity at $z = \pm \frac{1}{2g}$ whereas the brane solution
\eqref{gog} is everywhere non-singular. In the following section we examine 
in more detail the motion of a test particle in the two metrics from \eqref{gog} 
and \eqref{domain-wall} to see which metric most closely corresponds to a Newtonian
infinite mass plane in so far as giving a uniform acceleration. 

All the above metrics had $\delta -$ function sources (or, more correctly,
distributions). In linear theories such as electromagnetism or Newtonian gravitation these types of sources do not present a major problem. But the nonlinear nature of general relativity suggests that one should proceed with caution when dealing with concentrated sources. The formalism for treating thin shell distribution such as those
studied here can be found in \cite{israel}. However, the formalism for treating 
line and point distributions is more involved. An attempt to incorparate point distributions 
in general relativity can be found in \cite{goldberg}. A detailed study of the possible inclusion of point, line and string
distributional sources in general relativity can be found in \cite{geroch}. Starting from the notion of {\sl regular} metrics, which satisfy the physically reasonable requirement that the curvature tensor - and, through Einstein's equations, the energy-momentum tensor as well - make sense as a distribution, Geroch and Traschen show that point particles and strings are not allowed as sources. Thin shells of matter or radiation, on the other hand, do admit a general formulation as distributional sources under this requirement of regularity.

\section{Particle trajectories for infinite plane}

In this section we study the geodesic motion of a test particle in each of the 
background metrics \eqref{gog} and \eqref{domain-wall}. For our 
purposes the best form of the geodesic equations is given on page 330 of
reference \cite{ohanian} as

\begin{equation}
g_{\mu \nu }\frac{d^{2}x^{\nu }}{d\tau ^{2}}+\frac{\partial g_{\mu \nu }}{%
\partial x^{\alpha }}\frac{dx^{\alpha }}{d\tau }\frac{dx^{\nu }}{d\tau }-%
\frac{1}{2}\frac{\partial g_{\alpha \beta }}{\partial x^{\mu }}\frac{%
dx^{\alpha }}{d\tau }\frac{dx^{\beta }}{d\tau }=0.  \label{geo-general}
\end{equation}
We first apply this to the ``brane" metric of \eqref{gog} restricting
ourselves to $z>0$ so we drop the absolute value around $z$. The $\mu =0$ equation from \eqref{geo-general} is

\begin{equation}
\frac{d^{2}t}{d\tau ^{2}}+2g\frac{dz}{d\tau }\frac{dt}{d\tau }=0.
\end{equation}
This has the solution \cite{rohrlich} (we assume $x$ and $y$ are fixed; the consistency of this assumption may be verified directly from \eqref{geo-general})
\begin{equation}
\frac{dt}{d\tau }=e^{-2gz}  \label{dtau}
\end{equation}
From \eqref{geo-general} the geodesic equation for the $z$-dimension is

\begin{equation}
\frac{d^{2}z}{d\tau ^{2}}+\frac{1}{2}\frac{\partial (e^{2gz})}{\partial z}
\left( \frac{dt}{d\tau }\right) ^{2}-\frac{1}{2}\frac{\partial g_{ij}}{\partial z}
\frac{dx^{i}}{d\tau }\frac{dx^{j}}{d\tau } =0 ,  \label{z-geodesic}
\end{equation}
where $i,j$ ranges over $1,~2$. Substituting the result of \eqref{z-geodesic}, using
$\frac{\partial g_{ij}}{\partial z} = 2 g g_{ij} = 2 g e^{2gz} \eta_{ij}$ (where
$\eta _{ij} =+1$ if $i=j=1$ or $i=j=2$, and zero otherwise) we can write
the above geodesic equation as

\begin{equation}
\frac{d^{2}z}{d\tau ^{2}}=-ge^{-2gz}+ g e^{2gz} \eta_{ij} \frac{dx^{i}}{d\tau }
\frac{dx^{j}}{d\tau } ~, \label{prop-acc}
\end{equation}
The acceleration is not constant because of the $e^{-2gz}$
factor in the first term. This runs counter to a naive extension of Newtonian
intuition, as encapsulated in \eqref{laplace}, that the acceleration should
be constant. However, \eqref{prop-acc} is what is usually meant by {\it 4-acceleration 
in special relativity}: it uses the proper time $\tau$ of the test particle 
undergoing geodesic motion. The observer who measures this 
acceleration is fixed at some $z \ne 0$. In the context of special
relativity the acceleration defined in \eqref{prop-acc} is singled out
by being covariant. In the context of general relativity this acceleration
in \eqref{prop-acc} is not covariant and therefore does not
hold the same priviledged role that it does in special relativity. 
Thus, in order to see the connection with
the Newtonian case we follow \cite{desloge} and look at the acceleration  
(referred to as ``local" acceleration in \cite{desloge}) measured with a system of clocks 
that are fixed with respect to the infinite plane at $z=0$ (due to general
relativistic time dilation this observer is not the same as the previous
observer fixed at $z \ne 0$).
Time intervals $dT$ measured with these clocks are related to proper time intervals via 
$d\tau = \sqrt{- g_{00}} dT =\sqrt{e^{2gz}}dT$ \cite{desloge}. With
this re-parameterization the geodesic equation \eqref{prop-acc} for the 
$z$-dimension becomes

\begin{equation}
e^{-gz}\frac{d}{dT}\left( e^{-gz}\frac{dz}{dT}\right)
=-ge^{-2gz}+ g \eta_{ij}\frac{dx^{i}}{dT}\frac{dx^{j}}{dT}.
\label{geo-general2}
\end{equation}
This simplifies to
\begin{equation}
\label{ref-plane}
\frac{d^{2}z}{dT^{2}}=-g+g \left( v_z ^2 + e^{2gz} (v_x ^2 +v_y ^2) \right)
\end{equation}
where $v_{x}\equiv \frac{dx}{dT}$, $v_{y}\equiv \frac{dy}{dT}$ and $v_{z}\equiv \frac{dz}{dT}$.
Note that the acceleration in \eqref{ref-plane} is not covariant, even in 
the special relativistic sense. The first term in \eqref{ref-plane} gives the constant
acceleration toward the plane at $z=0$ which one would expect from the Newtonian
case. However, unlike the Newtonian case, both the 4-acceleration
\eqref{prop-acc} as well as the acceleration \eqref{ref-plane} measured with clocks fixed relative to the plane
have a dependence on the velocity, and do not therefore reproduce the Newtonian result in general. 

In \eqref{ref-plane} one can see that the velocities in the plane
(i.e. $v_x, v_y$) scale exponentially as one moves off of the plane
$z=0$. The same is true of momenta and some other physical quantities.
It is exactly this property of the metric \eqref{gog} that makes it useful
for addressing the hierarchy problem. In 5D it can be arranged so that 
the effective 4D Newton's constant, $G_4$, is related to the underlying
5D Newton's constant, $G_5$, via an exponential scaling, $G_4 \propto e^{-kz} G_5$,
where $k$ is some constant and $z$ is now the 5$^{th}$ dimension. In this
picture $G_5$ can be on the order of other couplings (electroweak, strong) and
$G_4$ is small because of the exponential suppression due to the metric. To actually 
arrange this one needs two planes -- one with $+\sigma$ and one with 
$-\sigma$. The details of this can be found in \cite{rubakov}.

We now study the motion of a test particle on the domain wall background 
of \eqref{domain-wall}. The $\mu =0$ component of \eqref{geo-general} is
\begin{equation}
\frac{d^{2}t}{d\tau ^{2}}+ \frac{g}{\sqrt{1 -2gz}} \frac{dz}{d\tau }\frac{dt}{d\tau }=0.
\end{equation}
This has the solution  
\begin{equation}
\frac{dt}{d\tau }= \left(1 -2 g z \right)^{1/2}  \label{dtau2}.
\end{equation}
Next, from \eqref{geo-general}, and using \eqref{dtau2}
the geodesic equation for the $z$-dimension is
\begin{equation}
\frac{d^{2}z}{d\tau ^{2}} = - \frac{1}{2}g - \frac{1}{2}g \left( 1 - 2 g z \right) ^{-1} 
\left( \frac{dz}{d \tau} \right) ^2 - g \left( 1 - 2 g z \right) ^{1/2} 
\eta_{ij} \frac{dx^{i}}{d\tau } \frac{dx^{j}}{d\tau } ,  
\label{z-geodesic2}
\end{equation}

The first term on the right hand side of \eqref{z-geodesic2} shows that the 
domain wall metric gives a uniform 4-acceleration of $g/2$ toward the plane. The
remaining terms on the right hand side show (as in the case of the
``brane" metric) that contrary to Newtonian intuition the 4-acceleration
has velocity dependent terms. Switching to time intervals measured with clocks fixed
with respect to the plane at $z=0$ via the transformation $d\tau = \sqrt{- g_{00}} dT =
( 1 - 2 g z ) ^{-1/4} dT$ we find

\begin{equation}
\label{ref-plane2}
\frac{d^{2}z}{dT^{2}}=-\frac{g}{2} \left( 1 - 2 g z \right) ^{-1/2} 
-g  \left( 1 - 2 g z \right) ^{1/2}\left( v_x ^2 +v_y ^2 \right) ~,
\end{equation}
where $v_{x}\equiv \frac{dx}{dT}$ and $v_{y}\equiv \frac{dy}{dT}$. Therefore the ``local" acceleration
is not constant, but diverges as one approaches $z=\frac{1}{2 g}$. 

To more quickly arrive at the acceleration for a given metric one may directly
use the definition for the initial, local
gravitational acceleration given in \cite{desloge} 
\begin{equation}
g(z)\equiv - \frac{1}{\sqrt{-g_{00}}}\left( \frac{d \sqrt{-g_{00}}}{dz}\right) .  \label{local-g}
\end{equation}
Applying \eqref{local-g} to the ``brane" world  \eqref{gog} and domain wall
metrics \eqref{domain-wall} yields
\begin{equation}
\label{local-acc}
g_{Brane}(z)=-g [ \Theta (z) - \Theta (-z) ] ~, \qquad  
g_{Domain-Wall}(z)=- \frac{g}{2  (1 -2 g |z|)} [ \Theta (z) - \Theta (-z) ]~.
\end{equation}
Thus using  \eqref{local-g} as our definition we find that it is the
brane world metric which gives a uniform acceleration, while the acceleration of the
domain wall diverges at $z=\frac{1}{2 g}$. This frame dependence of the acceleration,
which is not an issue in the case of the Newtonian infinite plane, is of course unavoidable
and expected in the general relativistic case.

\section{Discussion of the source terms and energy conditions}

From the previous discussion one finds that the two metrics put forward as
general relativisitic versions of the Newtonian infinite plane have energy-momentum
tensors (\eqref{em-tensor} and \eqref{em-tensor2}) which include not only 
mass-energy density terms (i.e. $T_{00}$) but have pressure or tension terms
(i.e. $T_{ii}$). In contrast, the matter source of the Newtonian plane is only from
a mass-energy density term. 

One can give a physical motivation for the appearance of pressures or tensions (i.e. negative
pressures) in the static, general relativistic solutions: a matter source with only
a plane of mass-energy -- $T_{00} \propto \delta (z)$ -- is not stable, but will collapse
under its gravitational self-attraction. To have a static configuration one must
stabilize the mass-energy density by pressures/tensions (and in the case of the
``brane" world metric by a cosmological constant). Further, these pressures/tensions play a significant 
role in the total gravitational field. In \cite{ipser} it was shown that in order for
an observer to remain stationary next to a plane with planar mass-energy density $\sigma$ 
and planar tensions $\tau$, they would have to accelerate {\it away} from
the plane  if $(\sigma - 2 \tau) >0$ and {\it toward} the plane if $(\sigma - 2 \tau) <0$.  
In either case the magnitude of the acceleration would be proportional
to $| \sigma - 2 \tau |$. In other words, tensions are gravitationally repulsive, while
pressures (negative tensions) are gravitationally attractive; positive mass-energy 
densities are gravitationally attractive, while negative mass-energy densities are
gravitationally repulsive. 

We can use this conclusion to further understand the results of the previous section
for the accelerations of particles that are near the plane. If one is near the plane $z=0$,
the acceleration for a test particle in the ``brane" world metric is
$ \simeq g$ and toward the plane (see \eqref{prop-acc} \eqref{ref-plane}), while
a test particle in the domain wall metric sees an acceleration of $\simeq g/2$ and
also toward the plane (see \eqref{z-geodesic2} \eqref{ref-plane2}). For the ``brane"
world sources in \eqref{em-tensor} we have $\sigma =  - \frac{g}{4\pi G}$ and
$\tau = - \frac{g}{4\pi G}$ (since tensions act as negative pressures). Thus 
for the ``brane" world sources  $(\sigma - 2 \tau) = \frac{g}{4\pi G} > 0$.
The domain wall sources have $\sigma =  \frac{g}{4\pi G}$ and $\tau = \frac{g}{16 \pi G}$
which yields $(\sigma - 2 \tau) = \frac{g}{8\pi G} > 0$. Thus both metrics give an
attractive acceleration toward $z=0$, but when one is near the plane the 
``brane" world acceleration is twice that of th domain wall acceleration. The reason 
for the acceleration toward the plane is different in the two cases: (i) for the ``brane" world
source the attraction due to the pressures dominates the repulsion from the 
negative mass-energy  density; (ii) For the domain wall source the attraction from
the positive mass-energy density dominates the repulsion coming from the tension. 

The unusual source terms for these plane solutions (especially the negative mass-energy 
density of the ``brane" world solution) provide a nice segue for the introduction of 
energy conditions. Three principal energy conditions are the weak energy condition
\begin{equation}
\label{weak}
T_{\mu \nu} V^\mu V^\nu \ge 0 \qquad \rightarrow \qquad  \rho \ge 0 ~~ \mathrm{and} ~~ 
\rho + p_i \ge 0 ~,
\end{equation} 
the strong energy condition
\begin{equation}
\label{strong}
\left( T_{\mu \nu} -\frac{T}{2} g_{\mu \nu} \right) V^\mu V^\nu \ge 0 
\qquad \rightarrow \qquad  \rho + p_i \ge 0 ~~~ \mathrm{and} ~~~ 
\rho + \sum _i p_i \ge 0 ~,
\end{equation}
and the dominant energy condition
\begin{equation}
\label{dom}
T_{\mu \nu} V^\mu V^\nu \ge 0 \qquad \mathrm{and} ~~ T_{\mu \nu} V^\nu ~~~
\mathrm {not ~~ timelike}
\qquad \rightarrow \qquad  \rho \ge 0 ~~~ \mathrm{and} ~~~ 
-\rho \ge p_i \ge \rho ~.
\end{equation}
In the above $V^\mu$ is a timelike vector and in the strong energy
condition $T = -T_{00} + \sum _i T_{ii}$. The first statement of each condition 
is given in terms of $T_{\mu \nu}$ while the second statement is given
in terms of densities and pressures -- $T_{00} \rightarrow \rho$ and
$T_{ii} \rightarrow p_i$. The dominant energy condition implies the weak condition
but not the strong condition.

The energy-momentum tensor of the domain wall \eqref{em-tensor2}
satisfies all three energy conditions. The ``brane" world sources \eqref{em-tensor} violate
the dominant and weak conditions yet satisfies the strong condition. Thus 
one might again prefer the domain wall metric over the ``brane" world 
metric, since it does not violate any of the energy conditions. Nevertheless
there are experimentally confirmed cases where all the above energy conditions
are violated. The prototypical example is the Casimir effect \cite{casimir}.
Further discussion about energy conditions and their
actual or potential violation can be found in \cite{visser2} \cite{thorne}.

\section{Summary and Conclusions}

In this article we have presented two metrics -- \eqref{gog} and \eqref{domain-wall} --
which can to some extent be considered as general relativistic versions of
the Newtonian infinite plane. Each has a planar mass-energy density -- 
$T_{00} \propto \delta (z)$ 
-- and each has (in some frame) an (initially) uniform acceleration
toward the plane. For the ``brane" world metric this is the frame fixed to the 
plane (see \eqref{ref-plane}) while for the domain wall solution it is the
proper frame (see \eqref{z-geodesic2}). There are many points in which these
general relativistic planes differ from the Newtonian plane: (i) both require pressures/tensions
in order to have a static solution; (ii) the ``brane" world solution in addition requires a
cosmological constant; (iii) the acceleration (in any frame) has velocity dependent terms
(iv) the acceleration is frame dependent. 

Although neither solution is new, both provide good examples for beginning general relativity 
students in terms of calculations and to illustrate conceptual as well as physical points
regarding general relativity. The metrics \eqref{gog} and \eqref{domain-wall} are
simple enough to provide a doable exercise of calculating the Einstein
tensor, $G_{\mu \nu}$. Both metrics provide a relatively straightforward application 
of the geodesic equations \eqref{geo-general}. There are additional projects or exercises that
one could assign to students that would help to give a deeper understanding of the physical 
meaning of the solutions: (i) In this paper we have computed accelerations toward the plane $z=0$ only.  One can have students probe the physical meaning of both metrics further by having them calculate the 
solutions to the geodesic equations \eqref{geo-general} in the $x$ or $y$ directions for 
each metric under the initial conditions $\frac{d x}{d \tau}=0$ and
$\frac{d y}{d \tau}=0$. They could then be asked to explain the apparent discrepancy between these solutions and the picture emerging from a consideration of (the Newtonian approximation to) tidal forces as given on page 41 of \cite{ohanian},
\begin{equation}
f^i = -m R^i{}_{0k0} \, x^k.
\end{equation} 
Since $R^x{}_{0x0}$ and $R^y{}_{0y0}$ are non-vanishing 
this clearly gives $f^x = f^y \neq 0$, implying the existence of forces that should make two test particles with initial separations $\Delta x_0 \neq 0$ and/or $\Delta y_0 \neq 0$ approach each other as they fall toward the plane. Such forces are, of course, completely absent in the case of the Newtonian plane.
(ii) It is apparent that the domain wall metric \eqref{domain-wall} has a horizon
at $z= \frac{1}{2g}$, but it is less obvious that the brane world metric \eqref{gog} 
also possesses a horizon at $z= \pm \infty$ where the metric becomes infinite. Even though the
distance from $z=\pm \infty$ to $z=0$ is infinite the proper time required to cover 
this interval is finite. The easiest way to see this is to alter the metric \eqref{gog}
by letting $g \rightarrow -g$ so that particles  are repelled from the $z=0$ plane and 
accelerate toward $z = \infty$. Solving the geodesic equation \eqref{z-geodesic} for the
initial conditions $v_y = v_x = 0$ and $z=0$ at $t=0$ yields \cite{rubakov}
\begin{equation}
\label{z-geo-soln}
z(T) = \frac{1}{2 g} \ln \left(1 + g^2 T^2 \right)
\end{equation}
The proper time for this particle can be determined via
\begin{equation}
\label{proper-time}
d \tau ^2 = - e^{-2 g |z(T)|} dT^2 + \left( \frac{d z(T)}{dT} \right)^2 dT^2
\end{equation}
Using \eqref{z-geo-soln} in \eqref{proper-time} and integrating gives a proper time
of $\tau = \frac{\pi}{2 g}$ for the particle to move from $z=0$ to $z= \infty$. Thus for either
an attracting brane ($+g$) or a repelling brane ($-g$) the particle will cover the infinite
distance in finite proper time. Thus even though the horizons are at infinity they are
nevertheless important in considerations of this metric.   
This also allows one to introduce a similiar feature for the Schwarzschild solution where
it takes an infinite time as seen by an outside stationary observer for a particle to fall through
the horizon at $r=2M$, but the proper time, as seen by an obsever who falls with the
particle, is finite. (iii) Instead of studying the motion of point particles in these
background metrics as was done in section III, one can instead investigate the wave equations
of various particles (spin $0, \frac{1}{2}, 1, 2$). For example, one could look at the
wave equation of a spin-$0$ particle for the brane world metric. This is done
by modifiying the Klein-Gordon equation for the spin-$0$ field, $\Phi$, as
\begin{equation}
\label{scalar-brane}
\frac{1}{\sqrt{-g}} \partial_\mu (\sqrt{-g} g^{\mu \nu} \partial_\nu
\Phi) + m_0^2 \Phi = 0~,
\end{equation}  
where $g$ is the determinant of the metric, and $m_0$ is the mass of the field
$\Phi$. The equation for a spin-$2$ tensor field is the same as that for the spin-$0$
field. The 4D wave equations for spin-$\frac{1}{2}$ and spin-$1$ fields are similiar.
A guide or hint for obtaining these wave equations in the brane world background is that
they can be found by reducing from the explicit 5D wave equations -- 
for spin $0,2$, see the second article in \cite{RS}, for spin $\frac{1}{2}$ see \cite {bajc}, 
and for spin $1$ see \cite{pomarol}. To simplify the problem, 
one should first try massless fields (i.e. set $m_0 = 0$ in wave equations like
\eqref{scalar-brane}). One should also use separation of variables of the fields with the
$x$ and $y$ directions being plane waves, e.g., in wave equations such as \eqref{scalar-brane},
try $\Phi (t, x,y,z) \propto e^{i(E t - p_x x - p_y y)} \psi (z)$ where $E$ is the energy of
the particle and $p_x, p_y$ are the $x, y$ momenta. In this way one obtains 
a differential equation for $\psi (z)$ and one can 
investigate if a particular spin field is trapped or confined near the plane at $z=0$
i.e. does the function $\psi (z)$ fall off as $z \rightarrow \infty$. In the 5D case
this gave rise to some unusual and still puzzling results: spin 0 and 2 fields are trapped
when one has a repelling brane $(-g)$ \cite{RS}; spin $\frac{1}{2}$ fields are trapped by an attracting
brane $(+g)$ \cite{bajc}; spin 1 fields are trapped by neither brane \cite{pomarol}. It would be
interesting to see if/how these results from 5D carry over into 4D and if it is possible to
reconcile the trapping or non-trapping behavior of the wave fields with the respective particle
results from the geodesic equations when the metric given in \eqref{gog} is attractive or repulsive
(obtained from the former by making the replacement $g \rightarrow -g$ in \eqref{gog}). 
(iv) In section III we studied the geodesic equations for a massive particle. One
could redo the analysis for a massless particle.  

With the 4D brane solution \eqref{gog} in hand one can also begin to discuss
(either via class lecture or through assigned exercises)
how the 5D versions reproduces effective 4D gravity at low energies and
how these solutions are used to address the heirarchy problem. 
Although the domain wall solutions are not phenomenologically viable (if they 
existed one would have seen evidence for them in, among other things, the
fluctuations of the cosmic microwave background), they nevertheless would
provide an introduction to other exotic cosmological solutions, such as
cosmic strings or monopoles, which are still possibilities and continue to be
experimentally sought.    
 
Conceptually, these examples may be used to show that the acceleration
will in general contain velocity terms, which is distinct from
the Newtonian case. In addition, one sees how different types of
energy-momentum gravitate -- positive mass-density and pressures lead to
gravitational attraction, while negative mass-density and tensions 
lead to repulsion. Finally,  one can introduce various energy conditions by examining the energy-momentum tensor for each solution. 

\begin{center}
\textbf{Acknowledgments}
\end{center}

D.S. would like to thank the CSU Fresno College of Science and Mathematics 
for a 2006, 2007 summer professional development grant. 

%%%%%%%%%%%%%%%%%%%%%%%%%%%%%%%%%%%%%%%%%%%


\begin{thebibliography}{99}

\bibitem{gron}  P.A. Amundsen and {\O }. Gr{\o }n, ``General static
plane-symmetric solutions of the Einstein-Maxwell equation'', Phys. Rev. D 
\textbf{27}, 1731-1739 (1983); {\O }. Gr{\o }n and E. Eriksen, ``Equivalence in two-,
three-, and four-dimensional space-times'', Int. J. Theo. Phys., \textbf{31}
, 1421-1432 (1992)

\bibitem{desloge}  E.A. Desloge, ``Nonequivalence of a uniformly
accelerating reference frame and a frame at rest in a uniform gravitational
field'', Am. J. Phys., \textbf{57}, 1121-1125 (1989)

\bibitem{Gog1}  M. Gogberashvili, ``Hierarchy problem in the shell universe
model'', Int. J. Mod. Phys., D\textbf{11}, 1635-1638 (2002); ``Gravitational
trapping for extended extra dimension'', ibid. 1639-1642 (2002).

\bibitem{Gog2}  M. Gogberashvili, ``Four dimensionality in noncompact
Kaluza-Klein model'', Mod. Phys. Lett. A\textbf{14}, 2025-2032 (1999);
ibid., ``Our world as an expanding shell'', Europhys. Lett. \textbf{49},
396-399 (2000)

\bibitem{RS}  L. Randall and R. Sundrum, ``A Large mass hierarchy from a
small extra dimension'', Phys. Rev. Lett. \textbf{83}, 3370-3373 (1999);
ibid., ``An alternative to compactification'', 4690-4693 (1999).

\bibitem{rubakov}  V.A. Rubakov, ``Large and infinite extra dimensions: An
Introduction'', Phys. Usp. \textbf{44}, 871-893 (2001)

\bibitem{taub} A.H. Taub, ``Isentropic hydrodynamics in plane symmetric 
spacetimes ", Phys. Rev. {\bf 103}, 454-467 (1956)

\bibitem{vilenkin} A. Vilenkin, ``Gravitational field of vacuum domain walls and
strings", Phys. Rev. {\bf D 23}, 852-857 (1981)

\bibitem{ipser} J. Ipser and P. Sikivie, ``Gravitationally
repulsive domain wall", Phys. Rev. {\bf D 30}, 712-719 (1984)

\bibitem{ajp-string} D. Marolf, ``Resource Letter NSST-1: The nature and status of string theory",
Am. J. Phys. {\bf 72}, 730-741 (2004) 

\bibitem{rindler}  W. Rindler, ``Kruskal Space and the Uniformly Accelerated
Frame'', Am. J. Phys., \textbf{34}, 1174-1178 (1966)

\bibitem{tilbrook}  D. Tilbrook, ``General coordinatisations of the flat
space-time of constant proper-acceleration'', Aust. J. Phys. \textbf{50},
851-868 (1997)

\bibitem{einstein11}  A. Einstein, ``On the influence of gravitation on the
propagation of light,'' Annals of Physics (Germany) \textbf{35}, 898 (1911).

\bibitem{visser}  M. Visser, ``An Exotic Class of Kaluza-Klein Models'',
Phys. Lett. B \textbf{159}, 22-25 (1985).

\bibitem{israel} W. Israel, ``Singular hypersurfaces and thin shells in general relativity" Nuovo Cimento {\bf 44B} 1-14, (1966). 
Correction in Nuovo Cimento {\bf 48B} 463, (1967).

\bibitem{goldberg} J. Goldberg, {\it Gravitation, An Introduction to Current Research}
edited by L. Witten (Wiley, New York 1962)

\bibitem{geroch} R. Geroch and J Traschen, Phys. Rev. {\bf D 36},``Strings and other 
distributional sources in general relativity" 
1017-1031 (1987)

\bibitem{ohanian}  H. C. Ohanian and R. Ruffini, \textit{Gravitation and Spacetime}, 2nd. ed. (W. W.
Norton \& Company, 1994).

\bibitem{rohrlich}  F. Rohrlich, ``The principle of equivalence,'' Annals of
Physics (New York), \textbf{22}, 169-191 (1963).

\bibitem{casimir} H.B.G Casimir, ``On the attraction between two perfectly 
conducting plates", Proc. Kon. Nederl. Akad. Wetenschap, {\bf 51}, 793-795
(1948)

\bibitem{visser2} M. Visser, \textit{Lorentzian Wormholes}, (AIP Press, New York,
1995), 115-136

\bibitem{thorne} M.S. Morris and K. Thorne, ``Wormholes in spacetime and their
use for interstellar travel: A tool for teaching general relativity", Am. J. Phys. 
{\bf 56}, 395-412 (1988)

\bibitem{bajc} B. Bajc and G. Gabadadze, Phys. Lett. {\bf B 474}, 
``Localization of matter and cosmological constant on a brane
in anti de Sitter space", 282--291 (2000).

\bibitem{pomarol} A. Pomarol, Phys. Lett. {\bf B 486}, 
``Gauge bosons in a five-dimensional theory with localized gravity", 153--157 (2000).

\end{thebibliography}
\end{document}